# Raman spectra of fine-grained materials from first principles

Maxim N. Popov [1✉], Jürgen Spitaler[1], Vignaswaran K. Veerapandiyan[1], Eric Bousquet [2], Jiri Hlinka [3] and Marco Deluca [1✉]

Raman spectroscopy is an advantageous method for studying the local structure of materials, but the interpretation of measured spectra is complicated by the presence of oblique phonons in polycrystals of polar materials. Whilst group theory considerations and standard ab initio calculations are helpful, they are often valid only for single crystals. In this paper, we introduce a method for computing Raman spectra of polycrystalline materials from first principles. We start from the standard approach based on the (Placzek) rotation invariants of the Raman tensors and extend it to include the effect of the coupling between the lattice vibrations and the induced electric field, and the electro-optic contribution, relevant for polar materials like ferroelectrics. As exemplified by applying the method to rhombohedral $BaTiO_3$, AlN, and $LiNbO_3$, such an extension brings the simulated Raman spectrum to a much better correspondence with the experimental one. Additional advantages of the method are that it is general, permits automation, and thus can be used in high-throughput fashion.

## INTRODUCTION

Raman spectroscopy is a powerful technique based on the inelastic scattering of light[1], which is being increasingly used as a tool for studying the local structure of materials. Its main advantages are higher spatial resolution compared to other laboratory-scale diffraction methods[2], relatively easy sample preparation, low cost, fast data acquisition, and the non-contact and non-destructive measurement capability that allows implementing in-situ and in-operando experimental methodologies. Raman spectra in crystals originate, mostly, from the interaction of lattice vibrations (phonons) with quanta of incident monochromatic radiation[1,3]. Each phonon mode is associated to a specific atomic vibration (for example, stretching or breathing vibrations of chemical bonds or groups of atoms) that is sensitive to any geometry perturbation such that the influence of strain, temperature, electric field, etc. can be determined by examining the Raman mode variations upon external stimuli or by compositional changes. Hence, Raman spectroscopy has proven very effective for studying the strain state[4], defects[5,6] and phase transitions[7] specifically in ceramics and semiconductors. The main difficulty when applying this technique and the obstacle to its further development, however, is the rather complicated interpretation of Raman spectra necessary for extracting relevant information about material properties. In the case of single crystals, the interpretation is greatly facilitated by group-theoretical considerations[3,8] supported by first principles calculations of lattice dynamics[9–16]. This procedure is possible because single crystals permit well-defined scattering geometries and the theory of Raman scattering in such materials is readily available. Unfortunately, it is not always possible to obtain single crystals. More often the materials are produced in form of a powder or a polycrystal, e.g., ceramics. In this case, the interpretation of the experiments is complicated by the fact that the observed Raman spectrum is, essentially, a superposition of the Raman spectra coming from several small crystallites (grains), each oriented in a different way, which often results in broad and asymmetric spectral features. Attempts to extract useful information from the Raman spectra collected from powders have been made already in the 70's. Burns and Scott[17,18] concluded that the peak maxima observed in the powder spectra match with those in the single crystals, illustrating their point by considering $PbTiO_3$ (PT) and Pb$(Zr,Ti)O_3$ (PZT) ceramics. Following Burns and Scott, Brya[19] argued that, in transparent polycrystals, it is also possible to extract information about the phonon mode symmetry. Later on, Gualberto and Argüello[20] have demonstrated by theoretical arguments and by taking $LiIO_3$ as an example, that the good agreement found by Burns and Scott when comparing the peak positions of the powder and single crystalline data is rather an exception than a rule. Apart from this early developments, there is also a first principles based method that is used for computing Raman spectra of polycrystals[21–24]. This method, however, is often unable to reproduce some key features of the observed spectra.

In this work, we propose a new procedure to calculate Raman spectra of polar, fine-grained materials from first principles, which allows to reproduce with unprecedented accuracy their asymmetrical and broad spectral shape, thereby enabling precise phonon mode assignment also in these materials. The method is based on the proper spherical averaging of a randomly-oriented grain/domain distribution as explained in the next section. We demonstrate the advantages of this method by applying it to the interpretation of Raman spectra of rhombohedral $BaTiO_3$, AlN, and $LiNbO_3$.

## RESULTS

### The spherical averaging method

In this section, we will elaborate on how Raman spectra can be calculated using ab initio methods. We begin with single crystals, as this knowledge forms the basis for understanding polycrystals. Then we proceed to the state-of-the-art method for computing Raman spectra of polycrystals from single-crystalline data and discuss its shortcomings as well as the origin of these shortcomings. Finally, we will discuss how they can be fixed by employing a proper averaging procedure, resulting in the spherical averaging method.

[1]Materials Center Leoben (MCL) Forschung GmbH, Roseggerstrasse 12, A-8700 Leoben, Austria. [2]Physique Théorique des Matériaux, QMAT, CESAM, Universite de Liège, B-4000 Sart-Tilman, Belgium. [3]Institute of Physics, Academy of Sciences of the Czech Republic, Na Slovance 2, 18221 Prague, Czech Republic. ✉email: maxim.popov@mcl.at; marco.deluca@mcl.at

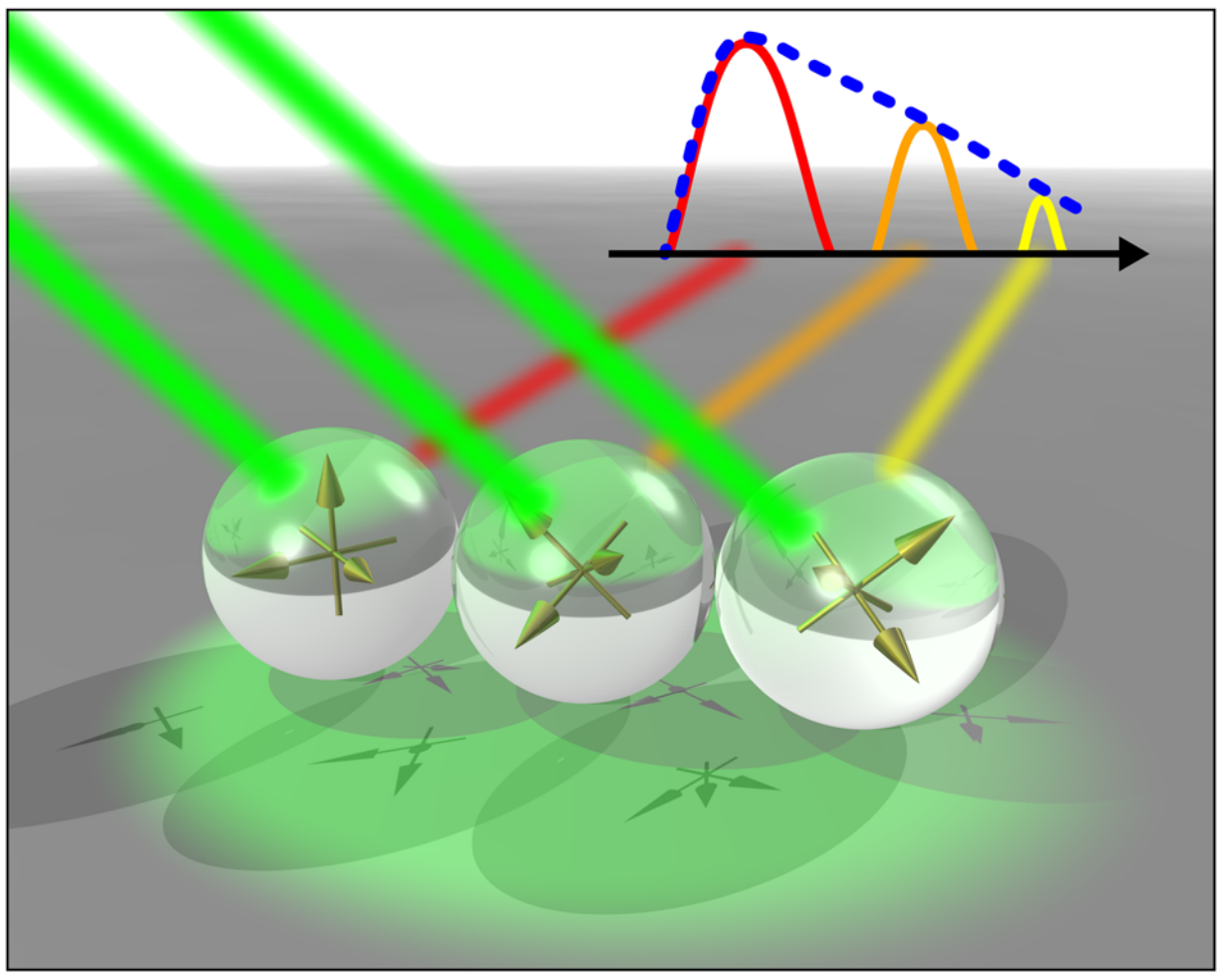

**Fig. 1 A schematic of Raman scattering on a polycrystal.** Each ball symbolizes a grain with its own orientation. The net spectrum is a superposition of the individual spectra coming from each grain.

Raman spectra originate from the interaction of incident light with phonons, rotons, magnons, and other quasiparticles in the matter, leading to the loss or gain of energy in the scattered light (inelastic scattering). This manifests itself in the shift of the light wavelength—the Raman shift with respect to the incident light wavelength. This shift corresponds to the energy of the quasiparticle involved in the scattering process, e.g., the phonon. That is, knowing the phonon energies/frequencies allows to identify the potential positions of the Raman shifts characteristic of the considered material. Whether the given peak will, indeed, appear in the measured spectrum and what would be its intensity if it does, depends on the material properties and the geometry of the scattering process. So, in single crystals, the peak intensity corresponding to a given vibrational mode (phonon) can be estimated as[25]

$$I \propto \left|\mathbf{e}^{s} \cdot \boldsymbol{\alpha} \cdot \mathbf{e}^{0}\right|^{2}, \tag{1}$$

where $\mathbf{e}^0$ and $\mathbf{e}^s$ are the polarisation vectors of the incoming and outgoing light (photons) and $\boldsymbol{\alpha}$ is the Raman susceptibility tensor (Raman tensor). The latter is defined as[25]:

$$\alpha_{ij} \propto \sum_{\kappa,\beta} \frac{\partial \chi^{(1)}_{ij}}{\partial \tau_{\kappa\beta}} u(\kappa\beta), \tag{1a}$$

where $\chi^{(1)}_{ij}$ is the linear dielectric susceptibility tensor at the frequency of the probing photon, $\tau_{\kappa\beta}$ is the shift of an atom $\kappa$ along the direction $\beta$, and $u(\kappa\beta)$ is the eigendisplacement of the same atom along the same direction, corresponding to the mode vibrational pattern. A superposition of the line shape functions, e.g., Lorentzian, centred at the mode frequencies ($\omega_i$) and scaled by their intensities ($I_i$) results in a simulated Raman spectrum $S(\omega)$:

$$S(\omega) = \sum_i I_i \cdot \frac{\sigma_i}{(\omega - \omega_i)^2 + \sigma_i^2}, \tag{2}$$

where $i$ is the mode index and $\sigma_i$ is the mode broadening, which depends on the phonon lifetime. Both phonon frequencies and Raman tensors can be computed from first-principles[9,11,25] using density-functional perturbation theory (DFPT). It is also possible to extract phonon frequencies and the Raman intensities from ab initio molecular dynamics simulations[26,27], which allow to capture anharmonic and some thermal effects. Since in crystalline materials the character of the excited vibration depends on the polarisation of incident and scattered light with respect to the orientation of the crystal, a thorough Raman mode assignment can be performed experimentally on single crystals by varying $\mathbf{e}^0$ and $\mathbf{e}^s$, and comparing the obtained spectra with the ones calculated from first principles and applying Eqs. (1) and (2).

Interpreting and predicting the Raman spectra of polycrystalline materials is a more elaborate task, as there is no clearly fixed experimental geometry relating the directions and polarisations of the incident and the scattered beams to the crystallographic directions of the probed material. Instead, the probed volume of the specimen constitutes an ensemble of differently-oriented grains (see Fig. 1). As a consequence, the scattered light is a composite response from many crystallites (grains)—each with its specific scattering geometry. Thus, the modelling of a Raman spectrum should involve averaging over this multitude of geometries. A common approach[21–24] resorts to averaging the mode Raman tensors of single crystals, yielding the so-called Placzek rotation invariants[1]:

$$G^{(0)} = \frac{1}{3}\left\{\left|\alpha_{xx} + \alpha_{yy} + \alpha_{zz}\right|^2\right\}, \tag{3}$$

$$G^{(1)} = \frac{1}{2}\left\{\left|\alpha_{xy} - \alpha_{yx}\right|^2 + \left|\alpha_{xz} - \alpha_{zx}\right|^2 + \left|\alpha_{yz} - \alpha_{zy}\right|^2\right\}, \tag{4}$$

$$\begin{aligned} G^{(2)} = {} & \tfrac{1}{2}\left\{\left|\alpha_{xy} + \alpha_{yx}\right|^2 + \left|\alpha_{xz} + \alpha_{zx}\right|^2 + \left|\alpha_{yz} + \alpha_{zy}\right|^2\right\} \\ & + \tfrac{1}{3}\left\{\left|\alpha_{xx} - \alpha_{yy}\right|^2 + \left|\alpha_{xx} - \alpha_{zz}\right|^2 + \left|\alpha_{yy} - \alpha_{zz}\right|^2\right\} \end{aligned} \tag{5}$$

These invariants are then combined to derive the average mode intensity[22–24]:

$$I^a_i = C \cdot \left(10G^{(0)}_i + 5G^{(1)}_i + 7G^{(2)}_i\right), \tag{6}$$

where C is a pre-factor that depends on the frequencies of the laser ($\omega_L$) and phonon ($\omega_i$), as well as on the temperature (via the Bose-Einstein function $n(\omega_i, T)$)[25,28]:

$$C \propto \frac{(\omega_L - \omega_i)^4}{\omega_i}\left[n(\omega_i, T) + 1\right]. \tag{6a}$$

To compute a simulated Raman spectrum of a polycrystal, $S_p(\omega)$, within the standard approach, one again resorts to Eq. (2), but now the mode intensities ($I_i$) are substituted by the average ones ($I^a_i$):

$$S_p(\omega) = \sum_i I^a_i \cdot \frac{\sigma_i}{(\omega - \omega_i)^2 + \sigma_i^2}. \tag{7}$$

Unfortunately, this approach fails to reproduce the essential features of the Raman spectra of polar polycrystalline materials. The reason for this failure is rooted in ignoring two key phenomena inherent to polar materials – the coupling of the polar lattice vibrations to the induced electric field and the electro-optic contribution. In general, the aforementioned coupling leads to a direction-dependent splitting between the longitudinal and transverse optical modes (LO/TO splitting)[3], and a directional dispersion of the phonon frequencies, which is accompanied by the mode mixing for any phonon propagation direction (**q**-vector) other than parallel or perpendicular to the optical axis/axes, also known as oblique phonons[29] (see Figs 2–4a). As a result, the Raman shift associated with a given phonon mode varies as function of the crystallite orientation. The second phenomenon, i.e., electro-optic contribution, affects the Raman tensor via altering the derivative of the dielectric susceptibility in the case of the LO phonons[25]:

$$\frac{\partial \chi^{(1)}_{ij}}{\partial \tau_{\kappa\lambda}} = \frac{\partial \chi^{(1)}_{ij}}{\partial \tau_{\kappa\lambda}}(\varepsilon = 0) - \frac{8\pi}{\Omega_0} \frac{\sum_l Z^*_{\kappa\lambda l} q_l}{\sum_{l,l'} q_l \varepsilon_{ll'} q_{l'}} \sum_l \chi^{(2)}_{ijl} q_l, \tag{8}$$

where $\frac{\partial \chi^{(1)}_{ij}}{\partial \tau_{\kappa\lambda}}(\varepsilon = 0)$ is the derivative assuming zero-field, $Z^*_{\kappa\lambda l}$ is the

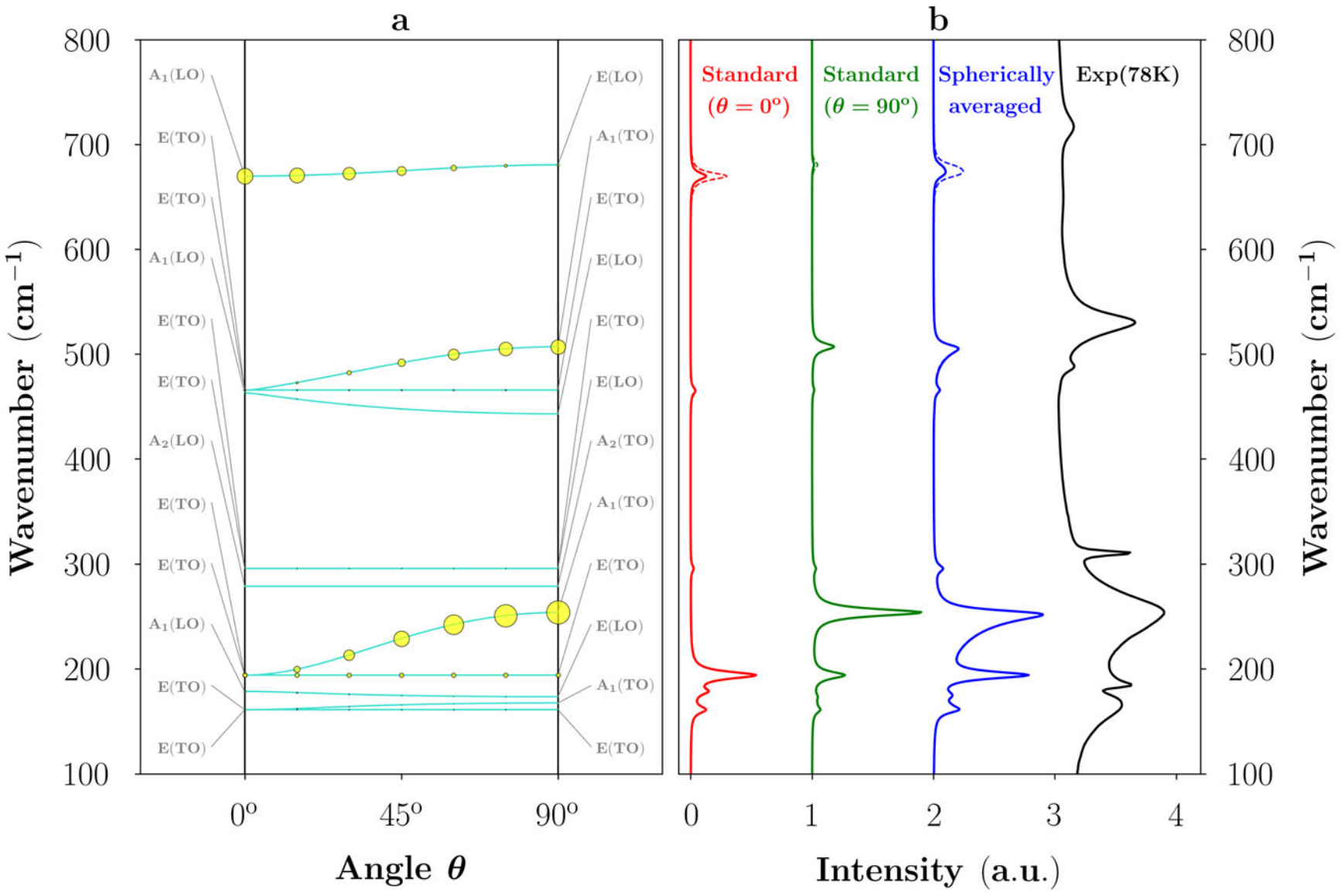

**Fig. 2 Phonons and Raman spectra of r-$BaTiO_3$.** **a** Computed angular dispersion of the phonon modes ($\theta$ is the angle between the $C_3$ axis and the phonon **q**-vector); the circle size is proportional to the mode Raman intensity, computed according to Eq. (6) and including the electro-optic contribution. **b** Simulated Raman spectra of polycrystalline r-$BaTiO_3$ compared to the experimental one. Dashed lines illustrate the effect of omitting the electro-optic contribution.

Born effective charge tensor, $q_l$ is the component of the phonon **q**-vector (taken in the $\mathbf{q} \to 0$ limit), and $\chi^{(2)}_{ijl}$ is the second-order nonlinear optical susceptibility. This additional contribution can be comparable to the zero-field derivative, moreover it is also clearly direction/orientation dependent. Neither the orientation-dependent peak shift caused by the LO/TO splitting, nor the direction-dependent electro-optic contribution are taken into account in the standard approach. At best, the LO/TO splitting and electro-optic term are computed for an arbitrarily chosen phonon propagation direction (**q**-vector). We will see in the next three sections how this arbitrariness results in quite different simulated Raman spectra and still does not capture the peak asymmetry. Hence, it is impossible to select the "best" **q**-vector to mimic the Raman spectrum of a polycrystal, instead, a proper averaging procedure is required. To devise such a procedure, one has to realise, first, that in polar materials Eq. (7) should, actually, read:

$$S_\mathrm{p}(\omega) = S_\mathrm{p}(\omega, \mathbf{q}) = \sum_i I_i^a(\mathbf{q}) \cdot \frac{\sigma_i}{(\omega - \omega_i(\mathbf{q}))^2 + \sigma_i^2}, \quad (9)$$

where **q** is the phonon wavevector (taken in the $\mathbf{q} \to 0$ limit). Next, we assert that this **q**-dependence should also be properly averaged out to yield the simulated Raman spectrum of a polar polycrystal or powder. Hence, we can express the averaged scattering intensity of a polycrystal $\overline{S_\mathrm{p}(\omega)}$ as:

$$\overline{S_\mathrm{p}(\omega)} = \frac{1}{4\pi} \iint d\theta d\varphi sin\theta S_\mathrm{p}(\omega, \theta, \varphi), \quad (10)$$

where $\theta$ and $\varphi$ are the components of the phonon **q**-vector in the spherical representation, and $S_\mathrm{p}(\omega, \theta, \varphi) = S_\mathrm{p}(\omega, \mathbf{q})$. The functions $S_\mathrm{p}(\omega, \mathbf{q})$ is calculated according to Eq. (9) using DFT and the integral can be evaluated numerically using the Lebedev-Laikov quadrature[30]. Equations (9) and (10), along with the reasoning behind them and the proposed integration scheme, constitute a new practical scheme for computing Raman spectra of polycrystalline materials, especially well-suited for polar materials, which we name the spherical averaging method. Since we do not make any assumptions about the symmetries of the considered material, our ab initio method is completely general, amenable to automation, and thus can be used in high-throughput fashion. The spherical averaging method can also be used in the case of non-polar materials, however, that will result in the calculated spectra being identical to those produced by just using the standard approach.

To demonstrate the performance of our method, we compute here the simulated Raman spectra of polycrystalline rhombohedral $BaTiO_3$, AlN, and $LiNbO_3$ and compare them to the experimental spectra, also obtained within this work. Raman spectra were always measured at 78–83 K (i.e., liquid $N_2$ temperature) to provide closer conditions to the simulated spectra (where $T = 0$ K).

## Simulated spectrum of rhombohedral-$BaTiO_3$

The first material we consider is perovskite $BaTiO_3$ in its low-T rhombohedral ferroelectric phase (R3m, space group no. 160), r-*$BaTiO_3$*. The unit cell of r-$BaTiO_3$ contains one formula unit, i.e., 5 atoms, yielding 12 optical modes. Factor group analysis of the $\Gamma$ point phonons predicts 11 Raman active modes ($3A_1 + 8E$) and 1 silent mode ($A_2$). These modes are directional, namely their frequency is dependent on their direction of propagation in the crystal or crystallite[19]. Pure $A_1$(LO,TO) and E(LO,TO) phonons exist only in the direction of the principal crystallographic axes. For all intermediate directions, only oblique phonons are visible with mixed frequencies between $A_1$(LO,TO) and E(LO,TO) modes[31]. Thus, essentially, in unpoled ferroelectric $BaTiO_3$ pure phonon frequencies are never measured but only oblique phonons can be obtained, regardless of polarisation direction. Only in poled ferroelectric crystals pure phonons can be estimated if the preferential (i.e. polar) crystal axis is known[32]. Given these characteristics of oblique phonons, no pure modes are visible in the polycrystalline spectrum of $BaTiO_3$. Nevertheless, previous studies assigned specific characters to the observable Raman peaks of rhombohedral $BaTiO_3$. The calculated and experimentally measured oblique phonons and the mode assignment obtained in this work are presented in Fig. 2a, in good agreement to literature. The three peaks at ~180–190 cm$^{-1}$ (the one in the middle is hardly visible) display E(LO/TO) and $A_1$(TO/LO) character; the broad and asymmetric peak at about 250 cm$^{-1}$ is mainly an $A_1$(TO) mode; the sharp peak at ~310 cm$^{-1}$ is due to the E(LO/TO) modes; the small peak at ~490 cm$^{-1}$ is a E(LO/TO) mode; the broad and asymmetric peak at about 550 cm$^{-1}$ is mainly an $A_1$(TO) mode mixed with E(TO), and finally the peak at ~720 cm$^{-1}$ is a combination of $A_1$ and E longitudinal phonon modes[33]. Most of these peaks present strong asymmetry due to the convolution of $A_1$/E peaks, LO/TO peaks and oblique phonon dispersion[7].

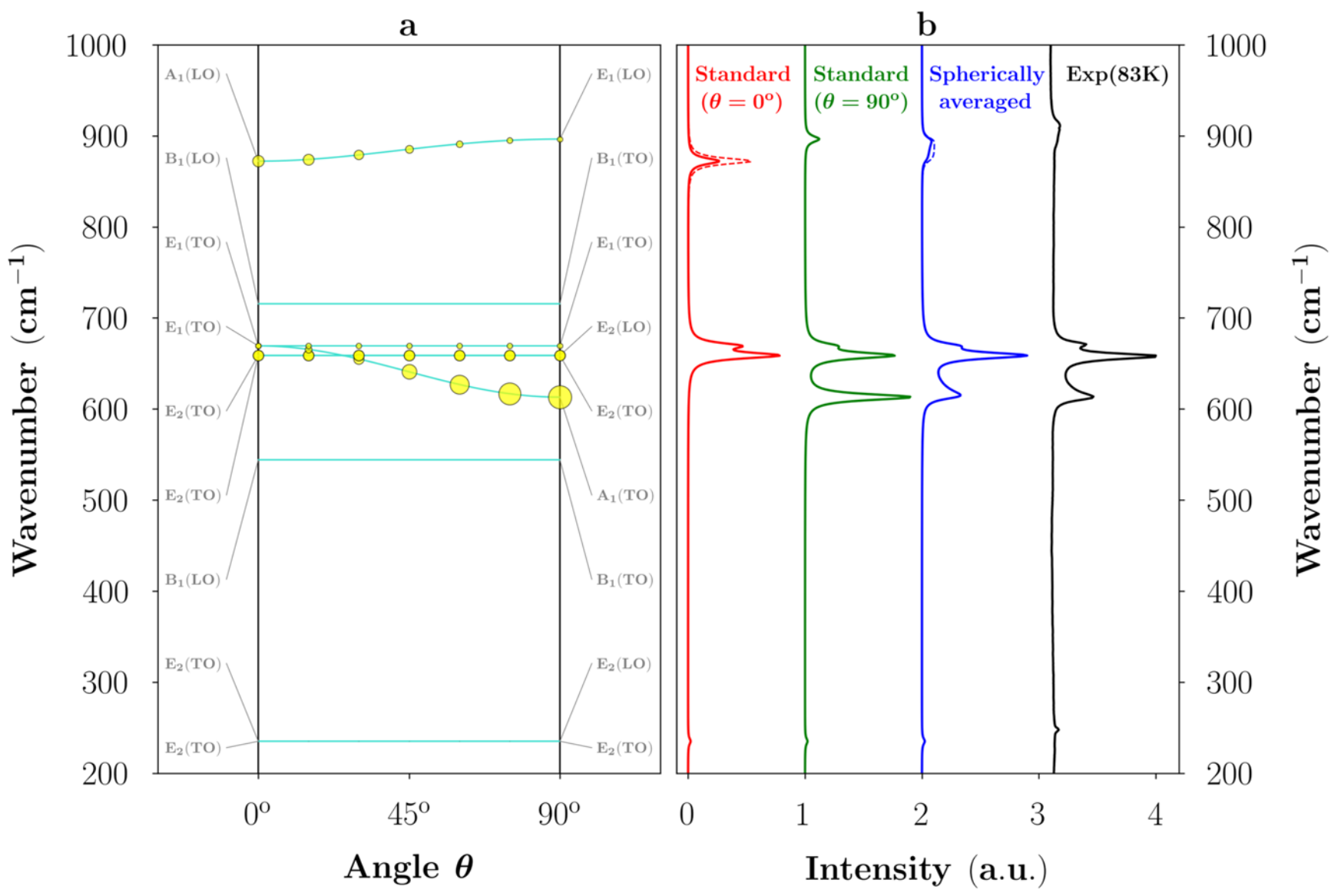

**Fig. 3 Phonons and Raman spectra of AlN. a** Computed angular dispersion of the phonon modes ($\theta$ is the angle between the $C_6$ axis and the phonon **q**-vector); the circle size is proportional to the mode Raman intensity, computed according to Eq. (6) and including the electro-optic contribution. **b** Simulated Raman spectra of polycrystalline AlN compared to the experimental one. Dashed lines illustrate the effect of omitting the electro-optic contribution.

Now, we will illustrate the ambiguity of the standard approach based on the rotation invariants of Raman tensors as well as its inability to reproduce the peak asymmetry. Figure 2b shows the Raman spectra computed using different approaches. The first two spectra (red and green) are obtained using the standard approach, choosing the **q**-vector either parallel ($\theta = 0°$) or perpendicular ($\theta = 90°$) to the $C_3$ symmetry axis. In the case **q** $\parallel$ $C_3$, the result shows very nice agreement with the theoretical spectra obtained by Hermet et al.[34] and ab-initio phonon dispersions previously published[35]. However, the overall appearance of the theoretical spectrum is quite different as compared to the experimental one (black line): the peak at 550 $cm^{-1}$ is absent and the relative peak intensities are wrong. When **q** is perpendicular to $C_3$, all peaks observed in the experiment are present and the relative intensities improve as well. However, the broad and asymmetric features at 250 and 550 $cm^{-1}$ are not captured. The latter is expected, since the simulated spectrum is a superposition of symmetric line-shape functions. Next, we consider the simulated spectrum computed using the spherical averaging method. Now, the simulated spectrum is in better agreement with the experiments, including the peak asymmetry. There are still discrepancies present, e.g., the under-estimated intensity of the lowest peak, and an overall shift in the peak positions towards lower values. Whereas the discrepancy on the intensities of the lowest peaks call for further investigation, the origin of the shift is likely to lie in the employed approximation for the exchange-correlation functional. As evident from Fig. 2, in addition, the electro-optic contribution is clearly important for obtaining the correct relative peak intensities (cf. solid/dashed blue lines).

## Simulated spectrum of AlN

As second example, we use wurtzite-structure aluminium nitride, AlN ($P6_3mc$, space group no. 186). Complications similar to those discussed for $BaTiO_3$ are also relevant for this material, namely LO/TO splitting, oblique phonon dispersion and convolution of different modes into broader features. This is evident from the Raman spectrum of fine powder AlN, which is shown in Fig. 3b (black line). The unit cell of AlN contains two formula units, i.e., 4 atoms, and hence there are 9 optical modes present. According to the group-theoretical analysis, there are 7 Raman active modes ($A_1+2E_1+4E_2$) and 2 silent modes ($2B_1$) in AlN. Within the considered wavenumber window there are four discernible peaks, namely at ~620 $cm^{-1}$ (asymmetric), ~660 $cm^{-1}$ (2 peaks), and ~910 $cm^{-1}$ (asymmetric). According to literature[36–39], the peak assignment is as follows: ~620 $cm^{-1}$ stems from $A_1$(TO), ~660 $cm^{-1}$ is composed of $E_2$ (main peak) and $E_1$(TO) (side peak of lower intensity at ~670 $cm^{-1}$) peaks, whereas the peak at ~910 $cm^{-1}$ is due to a combination of the $A_1$(LO) and $E_1$(LO) modes. The doubly-degenerate $E_2$-mode situated at ~235 $cm^{-1}$ is often disregarded in experimental spectra due to very low intensity. The computed oblique phonon dispersion along with the mode assignment performed within this work are presented in Fig. 3a. The mixing is evident in the case of the $A_1$-$E_1$ modes. The peak assignment agrees very well with the literature, except that the peak at ~620 $cm^{-1}$ is not only due to the $A_1$ mode, but also has a contribution from the $E_1$ mode (oblique phonon).

The simulated Raman spectra of powder AlN is presented in Fig. 3b. Again, we compare here the predictions of the standard approach, obtained for two exemplary **q**-points, to those given by the spherical averaging method. When the **q**-vector is chosen to be parallel to the $C_6$ symmetry axis, only three out of four peaks appear in the predicted spectrum. The prediction improves when the **q**-vector is perpendicular to $C_6$, however, the relative peak intensities are wrong and, moreover, the asymmetry of the peak at ~620 $cm^{-1}$ is not reproduced. When the spherical averaging method is used, the agreement to experiment is remarkable - both in shape and in the relative peak intensities. The electro-optic contribution only affects the highest-frequency mode and modifies its intensity and shape, leading to the best match to the experiment in the case of the spherical averaging method.

## Simulated spectrum of $LiNbO_3$

The last material we consider here is the ferroelectric phase of $LiNbO_3$. This material has a rhombohedral perovskite structure with space group R3c (no. 161). The unit cell of $LiNbO_3$ contains 10 atoms, giving rise to 27 optical modes. Group-theoretical analysis predicts 22 Raman active modes ($4A_1+18E$) and five silent modes ($5A_2$). The experimental spectrum collected from a powder sample is shown in Fig. 4b (black line). Overall, the spectrum exhibits a lot of broad peaks, whose origin can be understood by exploring the directional dispersion of the extraordinary (oblique) phonons in $LiNbO_3$[40]. The simulated Raman spectrum is given in Fig. 4a along with the mode assignment performed in this work, which agrees to the literature[40–42]. The shape of the predicted powder spectrum depends on the method used and again the best match to the experiment is achieved when applying the spherical averaging method. Also the relative peak intensities are reproduced very well in this case. The importance of the electro-optic contribution is evident in the case of highest-frequency mode (~840 cm$^{-1}$). As in the case of $BaTiO_3$, some peaks are shifted toward lower end. Again, we deem this to be mostly an effect of the chosen approximation for the exchange-correlation functional.

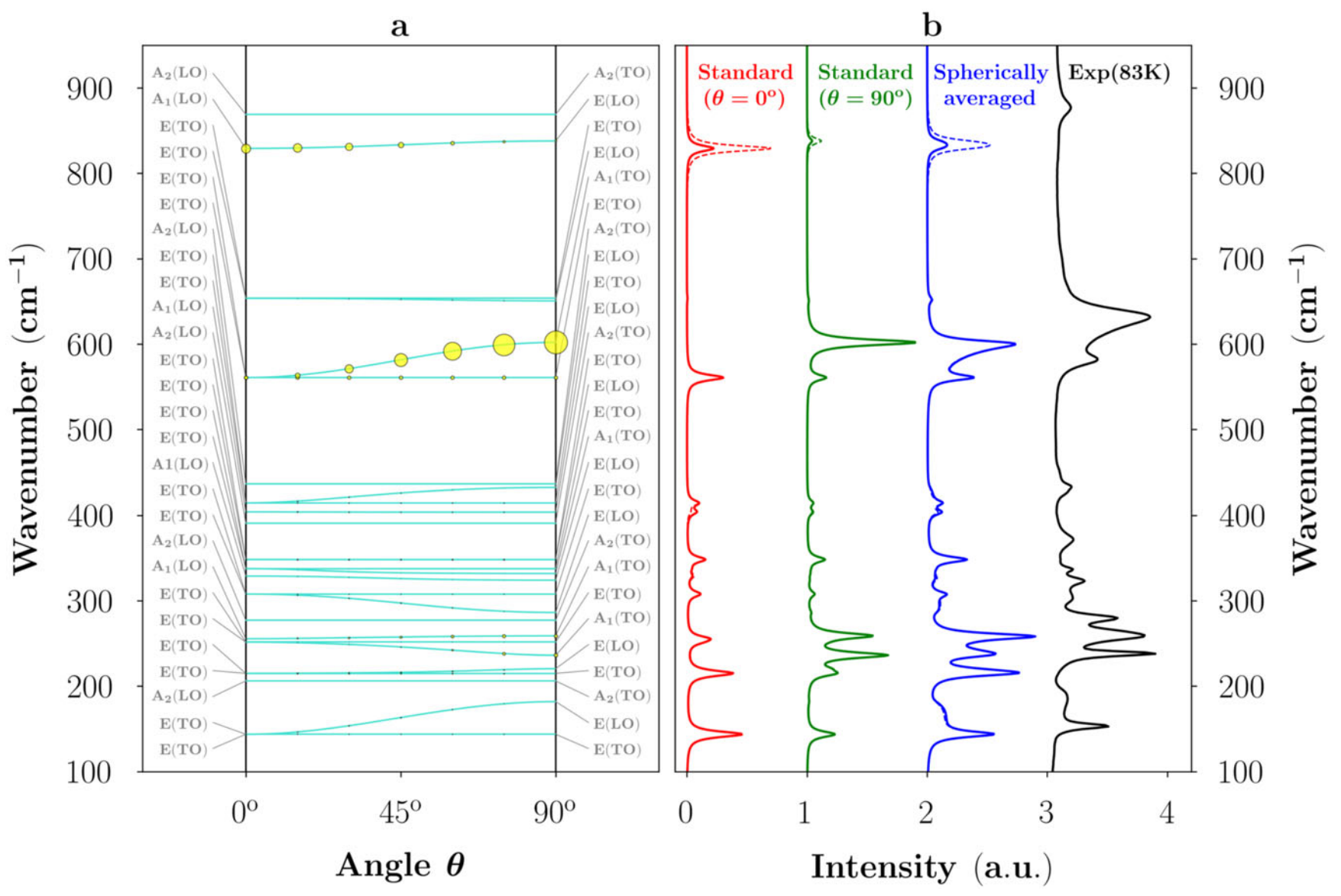

**Fig. 4 Phonons and Raman spectra of $LiNbO_3$. a** Computed angular dispersion of the phonon modes ($\theta$ is the angle between the $C_3$ axis and the phonon **q**-vector); the circle size is proportional to the mode Raman intensity, computed according to Eq. (6) and including the electro-optic contribution. **b** Simulated Raman spectra of polycrystalline $LiNbO_3$ compared to the experimental one. Dashed lines illustrate the effect of omitting the electro-optic contribution.

## DISCUSSION

In this paper, we address the aspects of predicting the Raman spectra of fine-grained materials (polycrystals, ceramics, or powders). In particular, we highlight the need for an additional averaging procedure in the case of polar materials, which takes into account the directional dependence of the phonon-frequencies and Raman tensors (in the **q** → 0 limit) caused by the coupling between the lattice vibrations and the induced electric field and electro-optic contribution. Based on this premise, we introduce a new ab initio method for computing simulated Raman spectra of polycrystalline, ceramic, or powder materials, which is suitable for high-throughput calculations. We validate the method by comparing the predicted spectra of rhombohedral perovskite $BaTiO_3$ and $LiNbO_3$, and wurtzite AlN, to the experimental ones—also obtained within this work on powder samples. It is confirmed that the new method allows to reproduce the experimental Raman spectra with unprecedented precision. Unlike the conventional approach based on the rotation invariants of Raman tensors, the new method allows to reproduce the broad and asymmetric peaks observed in the considered materials, and thus greatly improves the predictive power of ab-initio Raman spectra for any polar polycrystal. We are thus confident that our method will make an important contribution to the field of theoretical Raman spectroscopy of fine-grained materials.

## METHODS

### Modelling

For the calculation of Raman spectra, the density functional theory (DFT) was employed as implemented in the ABINIT code[25,43–45]. Before computing the simulated Raman spectra, all considered structures were fully relaxed (including the cell-shape) first using Wu-Cohen GGA[46], and then an additional ionic relaxation was made using LDA. The resulting structures were found to be in good agreement with the experimental and computational data from literature (see Supplementary Tables 1–4). After that, the phonons and Raman tensors were computed using LDA (due to constraints of ABINIT). The phonon frequencies and Raman tensors were determined within the density-functional perturbation theory (DFPT) as implemented in the ABINIT code[25,28]. The effect of temperature was only taken into account in terms of the phonon population statistics, while other effects like lattice expansion or temperature-dependent mode broadening, as well as anharmonicity, go beyond the scope of this work. Norm-conserving pseudopotentials were employed from the pseudodojo project[47] and regenerated for Wu-Cohen GGA exchange-correlation functional through the method and code developed by Hamann[48]. The energy cut-off for the wavefunction expansion was set equal to 55 Ha for all considered materials. To perform the integration over the Brillouin zone, the following k-meshes were used (all shifted): 8 × 8 × 8 for $BaTiO_3$, 8x8x4 for AlN, and 8 × 8 × 8 for $LiNbO_3$. While computing the simulated spectra, a uniform smearing width ($\sigma$) of 4 cm$^{-1}$ was applied to all modes and a dense Lebedev-Laikov mesh for spherical integration (consisting of 5810 points) was employed.

### Materials and experiments

High-purity fine-grained powders were used for Raman measurements of all investigated materials: $BaTiO_3$ (99.5% purity, Sigma-Aldrich, St. Louis, MO, USA); grade C AlN (abcr GmbH, Karlsruhe, Germany), and 99.9% pure $LiNbO_3$ (Sigma-Aldrich, St. Louis, MO, USA). The particle sizes of all powders were measured using the laser scattering-based CILAS 1064 L particle size analyser (CPS US Inc., Madison, WI, USA) in the range of 0.04–500 µm (wet method), giving the following values of $d_{50}$: $BaTiO_3$–1.38 µm; AlN–1.92 µm; $LiNbO_3$–21.05 µm. Details are reported in the Supplementary Information and visualised in supplementary Fig. 1.

Raman measurements were performed on a LabRAM 300 spectrometer (Horiba Jobin Yvon, Villeneuve d'Ascq, France) using an Nd:YAG solid state laser with a wavelength of 532 nm in a backscattering geometry. The laser light was focused on the sample surface by means of a long working distance 100x objective (with NA 0.8, LMPlan Fl, Olympus, Tokyo, Japan). The scattered radiation was dispersed with an 1800 gr/mm grating and visualised using a Peltier-cooled Charge Coupled Device (CCD). Linkam (THMS600, Linkam, Tadworth, UK) temperature-controlled stage was used for low T measurements at 78–83 K. The powders were compacted in small aluminium pans before placing them in the temperature stage to improve handling. Considering the aforementioned grain size of the powders and the fact that, in those materials, the Raman probe volume is about 600 $\mu m^3$ (cf. ref. [32]), spectra were collected on 15 or more different points per investigated material and averaged to ensure obtaining spectra representative of a polycrystalline material structure. The averaged spectra obtained in this way on each material (see supplementary figure 2) were then compared to the calculated ones.

## DATA AVAILABILITY

The data generated during this study are available from the corresponding author upon reasonable request.

## CODE AVAILABILITY

All code used to calculate the current results is available from the corresponding author upon reasonable request.

## ACKNOWLEDGEMENTS

M.N.P., J.S., V.K.V., and M.D. acknowledge financial support by the Austrian Science Fund (FWF): Project P29563-N36. This project has received funding also from the European Research Council (ERC) under the European Union's Horizon 2020 research and innovation programme (grant agreement No 817190). R.J. Bakker (Chair of Resource Mineralogy, Montanuniversitaet Leoben, Austria) is gratefully acknowledged for providing access to the Raman equipment. Prof. Franz Mautner, Brigitte Bitschnau (Institute of Physical and Theoretical Chemistry, Technical University of Graz, Austria) and Prof. Klaus Reichmann (Institute for Chemistry and Technology of Materials, Technical University of Graz, Austria) are gratefully acknowledged for their support with the characterisation of powders. EB thanks the FNRS, ARC AIMED project and the Consortium des Équipements de Calcul Intensif (CÉCI), funded by the F.R.S.-FNRS under Grant No. 2.5020.11 and Tier-1 supercomputer of the Fédération Wallonie-Bruxelles funded by the Walloon Region (Grant No. 1117545). J.H. was partially supported by Operational Program Research, Development and Education (financed by European Structural and Investment Funds and by the Czech Ministry of Education, Youth, and Sports), Project No. SOLID21-CZ.02.1.01/0.0/0.0/16_019/0000760). The computational results presented here have been achieved in part using the Vienna Scientific Cluster (VSC).

## AUTHOR CONTRIBUTIONS

M.N.P. has contributed to the methodological development, performed the DFT calculations and post- processing, took part in the discussion of the results, and prepared the initial draft of the manuscript. J.S. has contributed to the method development, discussing results, and providing calculation resource management. V.K.V. performed the Raman measurements and the related interpretation. E.B. provided support for DFT calculations using the ABINIT code. J.H. provided insights into the prior developments in the field of experimental Raman spectroscopy of powders. M.D. conceived and supervised the work, supported the methodological development providing insights about the features of the materials of study and the application of Raman spectroscopy, and led the projects on which this work is based. All authors contributed to preparing the manuscript.

## COMPETING INTERESTS

The authors declare no competing interests.

## ADDITIONAL INFORMATION

**Supplementary information** is available for this paper at https://doi.org/10.1038/s41524-020-00395-3.

**Correspondence** and requests for materials should be addressed to M.N.P. or M.D.

**Reprints and permission information** is available at http://www.nature.com/reprints

**Publisher's note** Springer Nature remains neutral with regard to jurisdictional claims in published maps and institutional affiliations.